\documentclass[aps,pre,showpacs,amsmath,amssymb]{revtex4}
\usepackage{graphicx,epsfig}
\usepackage{amsmath}
\usepackage{amsfonts}
\usepackage{amssymb}
\usepackage{mathptmx}
\usepackage{graphicx}

\begin{document}

\title{ON THE EMERGENCE OF COLLECTIVE ORDER IN SWARMING SYSTEMS: A
  RECENT DEBATE}

\author{M. Aldana}
\email{max@fis.unam.mx}

\author{H. Larralde}

\author{B. V\'azquez}
\affiliation{
Instituto de Ciencias F\'isicas,
Universidad Nacional Aut\'onoma de M\'exico.
Apartado Postal 48-3, C\'odigo Postal 62251, Cuernavaca, Morelos, M\'exico.}

\begin{abstract}
In this work we consider the phase transition from ordered to
disordered states that occur in the Vicsek model of self-propelled
particles. This model was proposed to describe the emergence of
collective order in swarming systems. When noise is added to the
motion of the particles, the onset of collective order occurs through
a dynamical phase transition. Based on their numerical results, Vicsek
and his colleagues originally concluded that this phase transition was of
second order (continuous). However, recent numerical evidence seems to
indicate that the phase transition might be of first order
(discontinuous), thus challenging Vicsek's original results. In this
work we review the evidence supporting both aspects of this debate. We
also show new numerical results indicating that the apparent
discontinuity of the phase transition may in fact be a numerical artifact
produced by the artificial periodicity of the boundary conditions.
\end{abstract}

\keywords{\it Swarming; Collective Motion; Phase Transition.}

\maketitle

\section{Introduction}

The collective motion of systems such as schools of fish, swarms of
insects, or flocks of birds, in which hundreds or thousands of
organisms move together in the same direction without the apparent
guidance of a leader, is one of the most spectacular examples of
large-scale organization observed in nature. From the physical point
of view, there has been a drive to try to determine and understand the
general principles that govern the emergence of collective order in
these systems, in which the interactions between individuals are
presumably of short-range. During the last 15 years, several models
have been proposed to account for the large-scale properties of flocks
and swarms. However, in spite of many efforts, the understanding of
these general principles has remained elusive, partly due to the fact
that the systems under study are not in equilibrium. Hence, the
standard theorems and techniques of statistical mechanics that work
well to explain the emergence of long range order in equilibrium
systems, cannot be applied to understand the occurrence of apparently
similar phenomena in large groups of moving organisms, or systems of
\emph{self-propelled particles} (SPP),as they have come to be known.  
\footnote{In this work we will refer to the individuals in the system
  as {\it particles}, other authors use instead the terms {\it agents}
  or {\it boids}.}

One of the simplest models conceived to describe swarming systems was
proposed by Vicsek and his collaborators in 1995, \cite{Vicsek-PRL-95}.
This model is ``simple'' in the sense that the interaction rules
between the particles are quite easy to formulate and implement in
computer simulations. However, it has been extremely difficult to
characterize its dynamical properties either analytically or
numerically. The central idea in Vicsek's model is that at every
time, each particle moves in the average direction of motion of its
neighbors, \emph{plus some noise}. In other words, each particle
follows its neighbors. But the particles make ``mistakes'' when
evaluating the direction of motion of their neighbors, and those
mistakes are the noise. There might be several sources of noise. For
instance, the environment can be blurred, and, consequently, the
particles do not see very well the direction in which their neighbors
are moving. We will call this type of noise \emph{extrinsic noise}
because it has to do with uncertainties in the particle-particle
``communication'' mechanism. On the other hand, it is possible to
imagine the particles as having some kind of ``free will'', so that
even when they know exactly the direction of their neighbors, they may
``decide'' to move in a different direction. We will call this last
type of noise the \emph{intrinsic noise} because it can be thought of
as being related to uncertainties in the internal decision-making
mechanism of the particles.

\begin{figure}[bt]
\centerline{\psfig{file=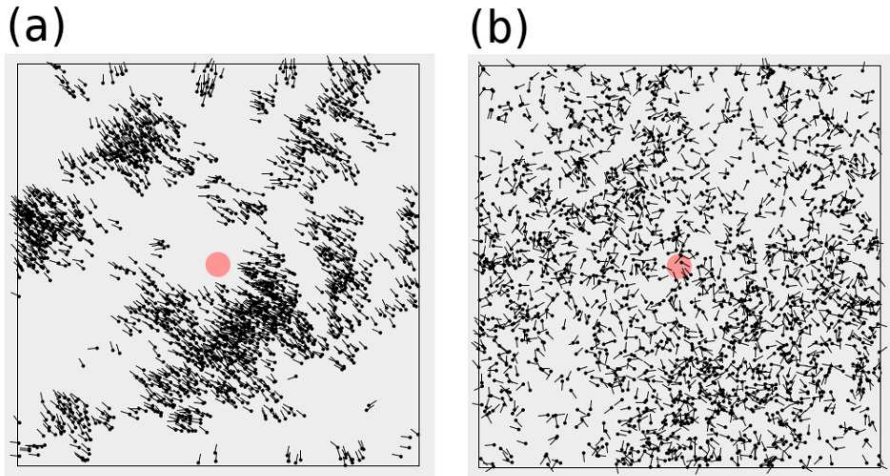,width=5in}}
\vspace*{8pt}
\caption{Snapshots of the Vicsek model. (a) For very small noise
  ($\eta = 0.1$ ) all the particles move in the same direction. (b)
  For very large noise ($\eta=0.8$) the particles move in random
  uncorrelated directions. The shaded circle at the center of each
  frame indicates the size of the interaction vicinity. In both cases
  intrinsic noise was used. The other parameters are: Number of
  particles, $N=2000$; size of the box, $L=32$; particle speed,
  $v_0=0.05$; interaction radius, $r_0=1$. }
\label{fig:1}
\end{figure}

In their original work, Vicsek et al. used intrinsic noise and showed
that, when the noise amplitude is increased, the system undergoes a
phase transition from an ordered state where all the particles move in
the same direction, to a disordered state where the particles move in
random uncorrelated directions (see Fig.\ref{fig:1}). Based on
numerical results, this phase transition was thought to be of second
order because it seemed to exhibit several of the characteristic
properties of critical phase transitions, such as a non-trivial
scaling behavior with the system size, divergence of the spatial
correlation length, etc. However, in 2004 Gr\'egoire and Chat\'e
challenged these results, \cite{Chate-PRL-04}. By numerically analyzing
large systems with many more particles than the ones studied by Vicsek
and his colleagues, Gr\'egoire and Chat\'e concluded that the phase
transition is in fact of first order for both types of noise,
intrinsic and extrinsic. This elicited a debate about the nature of
the phase transition in the Vicsek model, namely, whether this phase
transition is first order or second order. The existence of this
debate, into which some other research groups have got involved, is
indicative of the difficulties posed by the system, for not even such
a fundamental property as the nature of the phase transition occurring
in one of the simplest models of flocking, has been conclusively
resolved.

The question as to whether the phase transition is first order or
second order is not just of academic interest. Consider for instance a
school of fish. For low values of the noise, the nature of the phase
transition is irrelevant because the fish peacefully move in the same
direction in a dynamical state far away from the phase
transition. However, when the system is perturbed, for instance by a
predator, the entire school of fish collectively responds to the
attack: The school splits apart into smaller groups, which move in
different directions and merge again later, confusing the predator. A
possible hypothesis for this collective behavior to be possible is
that, when attacked, the fish adjust their internal parameters to put
the whole group close to the phase transition. If this phase
transition is of second order, then the spatial correlation length
diverges making it possible for the entire school to respond
collectively. However, this collective response would be very
difficult to mount if the phase transition was first order, since in
such a case the spatial correlation length typically remains finite.

In this work we review the evidence supporting the two sides to
this debate, i.e. the evidence supporting the continuous character
of the phase transition and the evidence supporting its discontinuous
character. It is important to mention that this is not a review on the
general topic of animal flocking. For such a review, we refer the
reader to Refs.~\cite{Couzin-Krauze-03}-\cite{Parrish-BB-02}. 

Here we focus only on the debate about the nature of the phase
transition in the Vicsek model \emph{with intrinsic 
noise}.\footnote{There is plenty of theoretical and numerical evidence
  showing that the phase transition in the \emph{extrinsic noise} case
  is indeed discontinuous. The problem is with the intrinsic noise.}
In the next section we introduce the mathematical formulation of this
model. In Sec.~\ref{sec:continuous} we present the work supporting the
conclusion that the phase transition is continuous, and in
Sec.~\ref{sec:discontinuous} the evidence supporting the opposite
conclusion. In Sec.~\ref{sec:artifact} we present new evidence
indicating that the discontinuous phase transition that has been
reported may be an artifact of the boundary conditions, which in the
numerical simulations performed so far have always been fully
periodic. Finally, in Sec.~\ref{sec:summary} we summarize the
different aspects of this debate.

\section{The Vicsek Model}
\label{sec:Vicsek}

The model consists of $N$ particles placed on a two dimensional square
box of sides $L$ with periodic boundary conditions. 
Each particle is characterized by its position
$\vec{x}_n(t)$, and its velocity $\vec{v}_n(t)=v_0 e^{i\theta_n(t)}$,
which we represent here as a complex number to emphasize the velocity
angle $\theta_n(t)$. All the particles move with the same speed
$v_0$. What changes from one particle's velocity to another is the
direction of motion given by the angle $\theta_n(t)$. In order to
state mathematically the interaction rule between the particles, let
us define $U_n(r_0;t)$ as the circular neighborhood or radius $r_0$
centered at $\vec{x}_n(t)$, and as $k_n(t)$ the number of particles
withing this neighborhood (including the $n$-th particle itself). Let
$\vec{V}_n(t)$ be the average velocity of the particles within the
neighborhood $U_n(r_0;t)$:
\begin{equation}
 \vec{V}_n(t) = \frac{1}{k_n(t)}\sum_{\{j:\vec{x}_j\in U_n\}} \vec{v}_j(t).
\label{eq:input}
\end{equation}

With these definitions, the interaction between the particles with
\emph{intrinsic noise} is given by the simultaneous updating of all
the velocity angles and all the particle positions according to the
rule
\begin{subequations}
\label{eqs:intrinsicRule}
\begin{eqnarray}
 \theta_n(t+\Delta t) &=&
 \mbox{Angle}\left[\vec{V}_n(t)\right]+\eta\xi_n(t), \label{eq:intTheta}\\ \vec{v}_n(t+\Delta
 t) &=& v_0 e^{i\theta_n(t+\Delta t)},\\ \vec{x}_n(t+\Delta t) &=&
 \vec{x}_n(t) + \vec{v}_n(t+\Delta t)\Delta t,
\end{eqnarray}
\end{subequations}
where the function ``Angle'' is defined in such a way that for every
vector $\vec{u}=ue^{i\theta}$ we have Angle$[\vec{u}] = \theta$. In
addition, $\xi_n(t)$ is a random variable uniformly distributed in the
interval $[-\pi,\pi]$, and the \emph{noise intensity} $\eta$ is a
positive parameter. Note that for $\eta=0$ the dynamics are fully
deterministic and the system reaches a perfectly ordered state,
whereas for $\eta \geq 1$ the particles undergo purely Brownian
motion.

Analogously, the interaction rule with \emph{extrinsic noise} is given by 
\begin{subequations}
\label{eqs:extrinsicRule}
\begin{eqnarray}
 \theta_n(t+\Delta t) &=& \mbox{Angle}\left[\vec{V}_n(t) + \eta e^{i\xi_n(t)}\right] \label{eq:extTheta}\\
 \vec{v}_n(t+\Delta t) &=& v_0 e^{i\theta_n(t+\Delta t)}\\
 \vec{x}_n(t+\Delta t) &=& \vec{x}_n(t) + \vec{v}_n(t+\Delta t)\Delta t
\end{eqnarray}
\end{subequations}
where $\xi_n(t)$ and $\eta$ have the same meaning as before. Note that
in Eq.~\eqref{eq:intTheta}, the angle of the average velocity
$\vec{V}_n(t)$ is computed and then a scalar noise $\eta\xi_n(t)$ is
added to this angle. On the other hand, in Eq.~\eqref{eq:extTheta} a
random vector $\eta e^{i\xi_n(t)}$ is added directly to the average
velocity $\vec{V}_n(t)$, and afterwards the angle of the resultant
vector is computed.\footnote{Some authors refer to the intrinsic noise
  $\eta\xi_n(t)$ as \emph{scalar noise}, and to the extrinsic noise
  $\eta e^{i\xi_n(t)}$ as \emph{vectorial noise}, \cite{Chate-PRL-04}.
  We prefer to use the terminology intrinsic and extrinsic,
  respectively, because we find it physically more appealing.}
Intuitively, we can interpret the average velocity $\vec{V}_n(t)$ as
the signal that the $n$-th particle receives from its neighborhood,
and the $Angle$ function as the particle's decision-making
mechanism. Therefore, in Eq.~\eqref{eq:intTheta} the particle clearly
receives the signal from its neighborhood, computes the angle of
motion, but then decides to move in a different direction. In
contrast, in Eq.~\eqref{eq:extTheta} the particle receives a noisy
signal from the neighborhood, but once the signal has been received,
the particle computes the perceived angle and follows it.

\begin{figure}[bt]
\centerline{\psfig{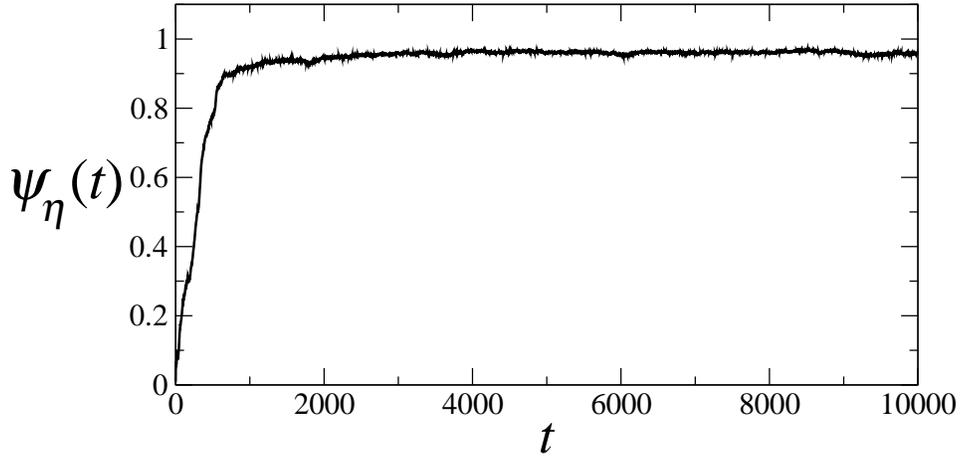}}
\vspace*{8pt}
\caption{Time series of the order parameter $\psi_\eta(t)$. Starting
  from a random initial condition for which $\psi_\eta(0)\approx 0$,
  after a transient time (about 2000 time steps in this case), the
  system reaches a steady state reflected by stationarity of the order
  parameter. The simulation was carried out using intrinsic noise for
  a system with $N=2\times10^4$, $\eta=0.1$, $v_0\Delta t = 0.05$ and
  $r_0 = 0.4$.}
\label{fig:2}
\end{figure}

At any time, the amount of order in the system is measured by the
instantaneous order parameter defined as
\begin{equation}
 \psi_\eta(t) = \frac{1}{N v_0}\left|\sum_{n=1}^{N}\vec{v}_n(t)\right|,
\label{eq:psi_t}
\end{equation}
where we have used the subscript $\eta$ to explicitly indicate that
the order in the system depends on the noise intensity. After a
transient time, the system reaches a steady state, as indicated in
Fig.~\ref{fig:2}. In the stationary state, we can average
$\psi_\eta(t)$ over noise realizations, or equivalently, over
time. The stationary order parameter $\psi(\eta)$ is, thus, defined as
\begin{equation}
 \psi(\eta) = \lim_{t\to\infty}\langle \psi_\eta(t)\rangle = \lim_{T\to\infty}\frac{1}{T}\int_0^T \psi_\eta(t)dt,
\label{eq:psi}
\end{equation}
where $\langle\cdot\rangle$ denotes the average over realizations,
whereas the average over time is given by the integral.

\begin{figure}[ht]
\centerline{\psfig{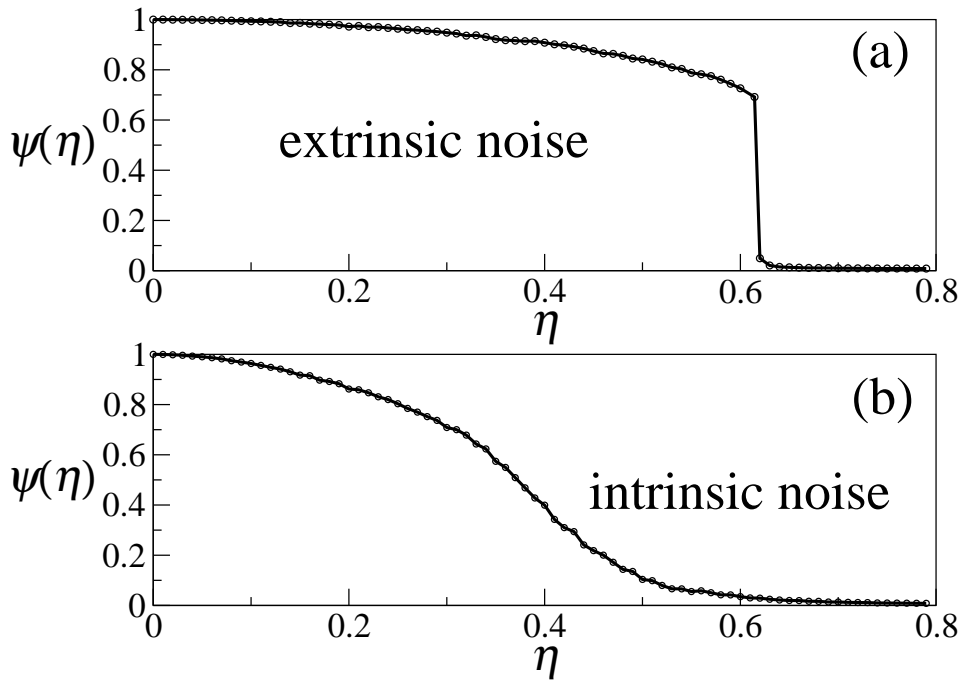}}
\vspace*{8pt}
\caption{Phase transition diagram for a system with (a) extrinsic
  noise and (b) intrinsic noise. All the other parameters are the same
  for both systems: $N=2\times10^4$, $L = 32$, $v_0\Delta t = 0.05$,
  and $r_0 = 0.4$. Note that in the extrinsic noise case the
  discontinuous phase transition is evident.}
\label{fig:3}
\end{figure}

Fig.~\ref{fig:3} shows typical phase transition curves for both
types of noise, intrinsic and extrinsic. The discontinuous character
of the phase transition for the extrinsic noise case is evident
(Fig.~\ref{fig:3}a). As far as we can tell, there are no doubts that
the phase transition is discontinuous in this case, and there is plenty
of numerical and theoretical evidence supporting this 
finding, \cite{Chate-PRL-04,Aldana-PRL-07,albano-ijmpc-06}.
In contrast, the phase transition shown in Fig.~\ref{fig:3}b for
the case with intrinsic noise seems to be continuous. However, despite
this apparent continuity, strong finite-size effects, reflected in the
smoothness of the curve as $\psi\to0$, may be hiding a possible
discontinuity, as has been claimed by Gr\'egoire and
Chat\'e, \cite{Chate-PRL-04}. For some as yet unknown reason, these
finite-size effects would be much stronger for the intrinsic noise
than for the extrinsic noise, masking one discontinuity but not the
other.

The numerical determination of the nature of a phase transition is a
difficult task precisely because the finite-size effects can conceal
any discontinuity either in the order parameter or in its
derivatives. Several techniques to deal with phase transitions
numerically have been developed
\cite{Lee-PRL-90}-\cite{Binder-RPP-97}. For the
particular case of the Vicsek model, and other closely related models,
finite-size scaling analysis and dynamic scaling
analysis \cite{Czirok-JPA-97}-\cite{Baglieto-PRE-08}, 
renormalization
group \cite{Toner-PRL-95}-\cite{Tonner-AP-05}, mean-field
theory \cite{Aldana-PRL-07}, \cite{Aldana-JSP-03}-\cite{Peruani-EPJ-08}, 
stability analysis \cite{Bertin-PRE-06},  and Binder
cumulants \cite{Chate-PRL-04,Chate-comment,Chate-PRE-08}, have been
used. However, none of these techniques have determined conclusively the
nature of the phase transition.

\begin{table}[bh]  
{\begin{tabular}{@{}c|c@{}} \hline 
\\[-1.8ex] 
Raw parameters & Effective parameters \\[0.8ex] 
\hline \\[-1.8ex] 
\begin{tabular}{@{}ll@{}}
$N$ & Number of particles \\ 
$L$ & Size of the box \\ 
$v_0$ & Particle speed \\ 
$r_0$ & Radius of interaction vicinity\\ 
$\Delta t$ & Integration time-step\\ 
$\eta$ & Noise intensity\\
{} & {} \\
{} & {} \\
\end{tabular} &
\begin{tabular}{@{}ll@{}}
$\rho=N/L^2$ & Density \\
$L'=L/r_0$ & Size of the box relative\\
{} & to the interaction radius\\
$K = \rho(\pi r_0^2)$ & Average number of \\
{} & interactions per particle \\
$l = (v_0\Delta t)/r_0 $ & Step size relative to\\
 {} & the interaction radius\\ 
$\eta$ & Noise intensity \\
\end{tabular}
\\[0.8ex] 
\hline 
\end{tabular}}
\caption{Parameters in the Vicsek Model}
\label{tab:parameters}
\end{table}

Another problem for determining the nature of the transition, in
addition to finite-size effects, is the existence of too many
parameters. This makes the numerical exploration of the parameter
space very difficult, and what is valid for one region may not be
valid for a different region. For instance, for low densities and
small particle speeds, the phase transition looks continuous for
systems as large as the computer capacity permits one to
analyze, \cite{Baglieto-PRE-08,Nagy-PA-07}. But for high densities,
large system sizes and high speeds the phase transition appears to be
discontinuous, \cite{Chate-PRL-04,Chate-PRE-08}.
Table~\ref{tab:parameters} lists all the parameters in the model. Of
particular importance is $l=(v_0\Delta t)/r_0$, the step size relative
to the interaction radius. When $l$ is small ($l<0.5$) we will say
that the system is in the low-velocity regime. Otherwise, the system
will be in the high-velocity regime.

\section{The Continuous Phase Transition}
\label{sec:continuous}

The first claim of the existence of a continuous phase transition in
the self-propelled particles model was presented in Vicsek's original
work \cite{Vicsek-PRL-95}, which was entitled ``Novel type of phase
transition in a system of self-driven particles.'' The adjective
``novel'' was used probably because this phase transition was not
expected in view of the Mermin-Wagner theorem \cite{Mermin-PRL-66},
which asserts that in a low-dimensional (1D and 2D) equilibrium system
with continuous symmetry and isotropic local interactions, there
cannot exist a long-range order-disorder phase
transition. Specifically, this theorem was proven having in mind
Heisenberg and XY models of ferromagnetism. Vicsek's model in 2D is
similar to the XY model, but in the latter the particles are fixed to
the nodes of a lattice, whereas in the former they are free to move
throughout the system. It is the motion of the particles that gives
rise to effective long-range spatial interactions among the particles
in the system. These interactions are responsible for the onset of
long-range order, producing the phase transition. We will come back to
this point later, (see Sec.\ref{sec:networks}).

\begin{figure}[t]
\centerline{\psfig{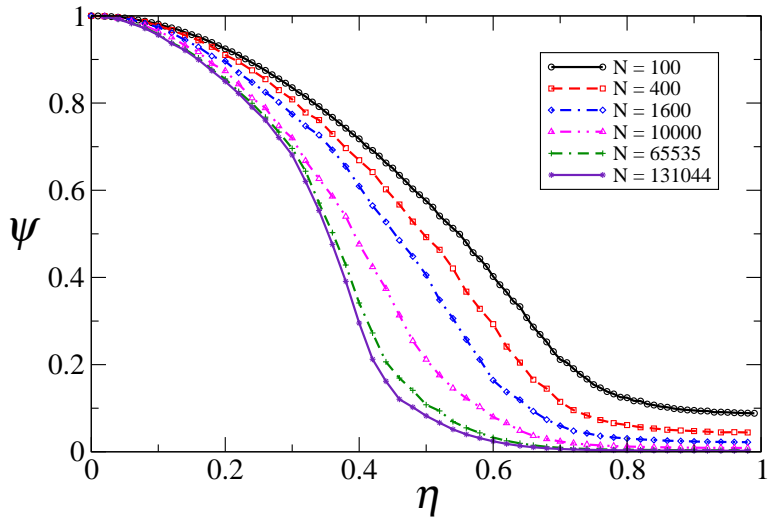}}
\vspace*{8pt}
\caption{Phase transition in the Vicsek model for different system
  sizes. In all the cases the following parameters were used:
  $\rho=4$, $r_0=1$, and $v_0\Delta t = 0.03$. Each point in each
  curve is the average of the instantaneous order parameter
  $\psi_\eta(t)$ over $10^6$ time steps.}
\label{fig:4}
\end{figure}

Fig.~\ref{fig:4} shows the phase transition for different system
sizes, keeping the density constant, $\rho = 4$. It should be
emphasized that the curves shown in this figure correspond to a
low-velocity regime in which the effective step-size $l=v_0\Delta
t/r_0$ is small: $l=0.03$. This regime may be biologically plausible
for systems such as flocks of birds, because birds can detect other
birds that are far away compared to the distance traveled by each
of them per unit of 
time.\footnote{Studies carried out for European
  Starlings (\emph{Sturnus vulgaris}) indicate that these particular
  birds interact with a fixed number of neighbors rather than with
  the ones within a given vicinity, \cite{Ballerini-PNAS-08}.} 
In Fig.~\ref{fig:4}, which is similar to the one originally
published in Ref.~\cite{Vicsek-PRL-95} but with more particles,
one can observe that the phase transition remains continuous even for
very large systems, up to $N=131044$. It is important to analyze very
large systems because, as noted by Gr\'egoire and
Chat\'e \cite{Chate-PRL-04,Chate-comment,Chate-PRE-08}, phase
transitions occur in the thermodynamic limit defined as $N\to\infty$,
$L\to\infty$, $\rho=const$. Using systems with different number of
particles, up to $N=10^5$, Vicsek and his collaborators found
numerically that the order parameter $\psi$ goes continuously to zero
when the noise amplitude $\eta$ approaches a critical value $\eta_c$
from below as
\begin{equation}
 \psi\approx(\eta_c - \eta)^\beta,
\label{eq:psi-beta}
\end{equation}
with a non-trivial critical exponent
$\beta=0.42\pm0.03$, \cite{Czirok-JPA-97,Czirok-PA-00}. The critical
value $\eta_c$ of the noise at which the phase transition seems to
occur depends on the system size $L$ and on the density $\rho$.  In
this situation, finite-size scaling analysis can be done to check
whether Eq.~(\ref{eq:psi-beta}) can be expected to be valid in the
thermodynamic limit, and to determine if a universal function
$\tilde\psi(\eta/\eta_c)$ exists such that the relationship
$\psi(\eta) = \tilde\psi(\eta/\eta_c)$ holds for any size $L$ and
density $\rho$.

\subsection{Finite-size scaling analysis}

\begin{figure}[t]
\centerline{\psfig{file=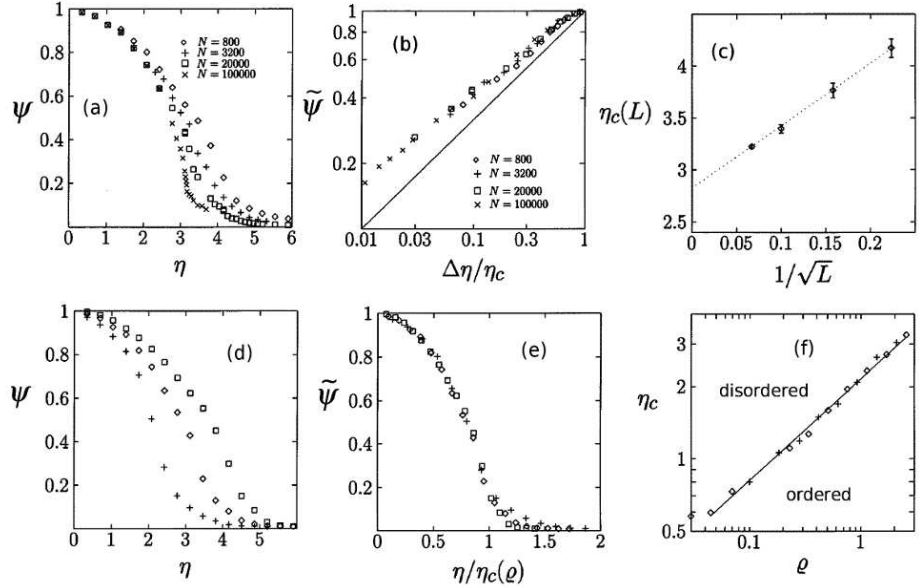,width=5in}}
\vspace*{8pt}
\caption{Scaling of the phase transition with the system size, [panels
    (a), (b) and (c)], and with the density $\rho$, [panels (d), (e)
    and (f)]. In the three top panels the density was kept constant,
  $\rho = 2$, and the number of particles was increased. In the three
  bottom panels the system size was kept constant, $L=100$, and the
  density varied, ($\rho = 4$ squares, $\rho = 2$ diamonds, $\rho =
  0.5$ pluses). In all the cases the system was in the low-velocity
  regime $l = 0.1$. The figures were taken from Ref.~\cite{Czirok-JPA-97}. In that
  work the authors used $\eta\in[0,2\pi]$ and $\xi\in[-0.5,0.5]$,
  whereas here we use $\eta\in[0 , 1]$ and $\xi\in[-\pi,\pi]$. In (b)
  $\Delta \eta = \eta_c - \eta$. }
\label{fig:5}
\end{figure}

In Refs.~\cite{Czirok-JPA-97,Czirok-PA-00}, Vicsek and his group
presented a finite-size scaling analysis showing that a universal
function $\tilde{\psi}(\eta/\eta_c)$ does indeed exist in the
low-velocity regime. Fig.~\ref{fig:5} shows how the phase
transition scales with the system size $L$ and with the density
$\rho$. From Figs.~\ref{fig:5}b and \ref{fig:5}e it is
clear that the order parameter computed for different system sizes and
different densities, plotted as a function of $\eta/\eta_c$, falls
onto the same universal curve $\tilde{\psi}(\eta/\eta_c)$.
Additionally, the critical value of the noise $\eta_c$ scales with the
system size and the density respectively as $\eta_c\approx L^{-1/2}$,
and $\eta_c\approx \rho^{\kappa}$, with $\kappa\approx 0.45$. A
similar scaling behavior was found for systems in 1D and
3D \cite{Czirok-PA-99,Czirok-PRL-99}. It should be noted that the
scaling behavior reported in Fig.~\ref{fig:5} is consistent with
of second-order phase transitions.

More recently, Baglieto and Albano performed a more thorough
analysis \cite{Baglieto-PRE-08}, finding consistent evidence of a
second order phase transition in the low-velocity regime (with
$l=0.1$) by means of two independent approaches. The first approach is
finite-size scaling analysis, in which they measure stationary
properties of the system such as the order parameter $\psi(\eta)$ and
the susceptibility $\chi=\langle\psi^2\rangle -
\langle\psi\rangle^2$. They assume that the following scaling
relations, which are well established in equilibrium systems, hold for
the Vicsek model close to the phase transition:
\begin{eqnarray}
 \psi(\eta,L) &\approx& L^{-\beta/\nu}\tilde{\psi}\left((\eta_c
 -\eta)L^{1/\nu}\right), \label{eq:albano1}\\ \chi(\eta,L) &\approx&
 L^{\gamma/\nu}\tilde{\chi}\left((\eta_c -\eta)L^{1/\nu}\right).
\end{eqnarray}
where $\tilde{\psi}$ and $\tilde{\chi}$ are the (universal) scaling functions.  

The second approach consists in performing short-time dynamic
simulations and studying the results on the basis of dynamic-scaling
theory, \cite{Zheng-IJMPB-98}. The main idea is to start out the
dynamics from fully ordered ($\psi_\eta(t) = 1$) and fully disordered
($\psi_\eta(t) \approx 0$) configurations, and analyze how the
system relaxes in time to the steady state. Close to the phase
transition, the relaxation from an ordered state is characterized by
the scaling relations \cite{Zheng-IJMPB-98}
\begin{eqnarray}
 \psi_\eta(t,L) &\approx&
 b^{-\beta/\nu}\hat{\psi}\left(b^{-z}t,b^\nu(\eta_c -\eta),
 b^{-1}L\right), \\ \chi_\eta(t,L) &\approx&
 b^{\gamma/\nu}\hat{\chi}\left(b^{-z}t,b^\nu(\eta_c -\eta),
 b^{-1}L\right). \label{eq:albano2}
\end{eqnarray}
where $z$ is the dynamic exponent, $b$ is an arbitrary scale factor,
and $\hat{\psi}$ and $\hat{\chi}$ are the universal dynamic scaling
functions. Additionally, it is known for equilibrium critical
phenomena that the so-called hyperscaling relationship
\begin{equation}
 d\nu-2\beta = \gamma,
 \label{eq:albano3}
\end{equation}
must be satisfied \cite{Binney-92}. Baglieto and Albano showed
numerically the validity of Eqs.(\ref{eq:albano1})-(\ref{eq:albano3})
for systems of different sizes and densities, obtaining excellent
agreement between the numerical results and the expected scaling
relationships. Thus, when finite-size effects are taken into account
in the Vicsek model with intrinsic noise and in the low-velocity
regime, two independent methods, one static and the other dynamic,
show consistency with a second-order phase transition.

\subsection{Mean-field theory and network models}
\label{sec:networks}

\begin{figure}[t]
\centerline{\psfig{file=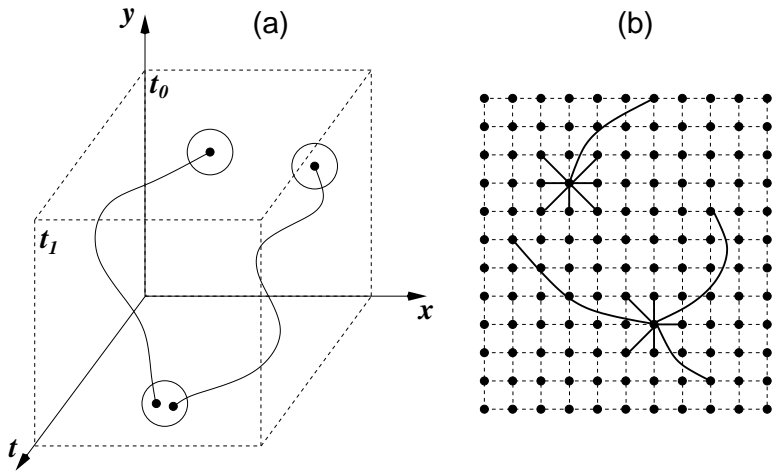,width=4in}}
\vspace*{8pt}
\caption{(a) Schematic representation of the long-range spatial
  correlations developed in time in the Vicsek model due to the motion
  of the particles. (b) A network model in which long-range
  interactions have been explicitly introduced via a small-world
  topology.}
\label{fig:6}
\end{figure}

As mentioned above, the XY model is similar to the Vicsek model in
that neighboring particles tend to align their ``spins'' (or
velocities) in the same direction. However, it is well known that in
the XY model, in which the particles do not move but are fixed to the
nodes of a lattice, there is no long-range order-disorder phase
transition \cite{Binney-92}. Thus, it is the motion of the particles in
Vicsek's model that enables the system to undergo a phase
transition. What this motion actually does is to generate throughout
time long-range spatial correlations between the particles. This is
illustrated in Fig.~\ref{fig:6}a, in which two particles that at
time $t_0$ do not interact, move and come together within the
interaction radius at a later time $t_1$. In
Refs.~\cite{Aldana-PRL-07} and \cite{Aldana-JSP-03} a
network model that attempts to capture this behavior is proposed. In
this model the particles are placed on the nodes of a small-world
network \cite{small-world}, in which each particle has $K$
connections. The small-world property consists in that a fraction $p$
of these connections are chosen randomly whereas the other fraction
$1-p$ are nearest-neighbor connections (see
Fig.~\ref{fig:6}b). As in Vicsek's and the XY models, each
particle is characterized by a 2D vector (or ``spin''),
$\vec{v}_n(t)=v_0e^{i\theta_n(t)}$, of constant magnitude and whose
direction changes in time according to Eqs.~(\ref{eqs:intrinsicRule})
and (\ref{eq:input}). However, in the network model the sum appearing
in Eq.~(\ref{eq:input}) is replace by the sum over the $K$ nodes
connected to $\vec{v}_n$.

\begin{figure}[t]
\centerline{\psfig{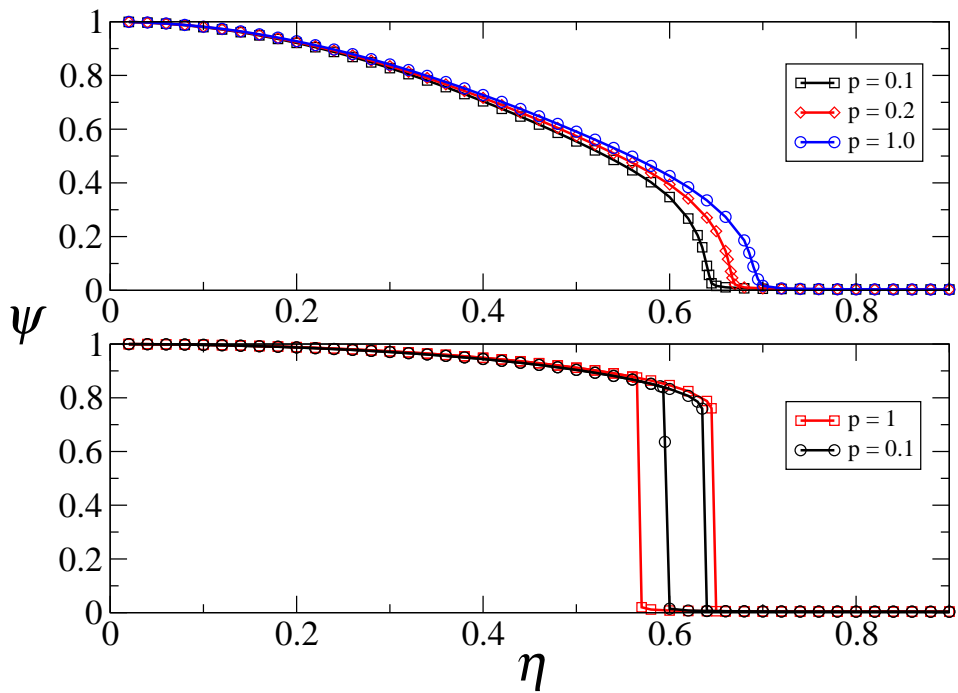}}
\vspace*{8pt}
\caption{Phase transition in the network model with (a) intrinsic
  noise and (b) extrinsic noise. Note that in the first case the phase
  transition is continuous whereas in the second case it is
  discontinuous with a region of hysteresis. The network models can be
  considered as the mean-field description of the Vicsek model,
  especially for $p=1$. However, the same behavior is observed for any
  non-zero value of $p$, even for very small ones. For $p=0$ the small
  world network model is analogous to the  XY model and
  no long-range order phase transition exists.}
\label{fig:7}
\end{figure}

The network model described above differs from the Vicsek model in
that the particles do not move, but are fixed to the nodes of a
network. However, the long-range spatial correlations created by the
motion of the particles in Vicsek's model are mimicked in the network
by the small-world topology, namely, by the $pKN$ random connections
scattered throughout the system. Although the analogy is not exact,
the network model can be considered as the mean-field version of
Vicsek's model, particularly when $p=1$, i.e. when the $K$ connections
of each particle come randomly from anywhere in the system. In this case
all spatial correlations are lost and the mean-field theory is
exact. In Ref.~\cite{Aldana-JSP-03} it is shown analytically that
the network model with intrinsic noise and $p=1$ undergoes a
second-order phase transition. In contrast, in
Refs.~\cite{Aldana-PRL-07,Pimentel-PRE-08} it is proved also
analytically, and for $p=1$, that the network model with extrinsic
noise exhibits a first-order phase transition. In Ref.~\cite{Aldana-PRL-07} 
it was shown numerically
that the network model with $p=1$ corresponds to the Vicsek model in
the limiting case of very large particle speeds ($l\gg1$). This is
because of the fact that for very large particle speeds, two
neighboring particles that at time $t_0$ are within the same
interaction vicinity, because of the noise in the direction of motion
of each particle, at the next time step $t_0+\Delta t$, they will end up
far away from each other and interact with other particles. Thus, in
the limit $v_0\to\infty$ of the Vicsek model, all spatial correlations
are destroyed in one single time step, which is exactly the condition
for the mean-field theory to be exact.

Numerical simulations show that for any $p\in(0,1]$, the phase
transitions with intrinsic noise is continuous, whereas it is
discontinuous with extrinsic noise (see
Fig.~\ref{fig:7}). The results obtained with the network model
indicate that the way in which the noise is introduced into the system
can indeed change the nature of the phase transition. Therefore, the
fact that the extrinsic noise produces a first-order transition cannot
be used as supporting evidence for the continuous or discontinuous
character of the transition with intrinsic 
noise.\footnote{In Ref.~\cite{Chate-PRL-04} the authors never claim
  that the results obtained for one type of noise extrapolate to the
  other type of noise. However, several authors interpreted the clear
  discontinuity of the phase transition with extrinsic noise as
  supporting evidence for the claim that the phase transition is also
  discontinuous in the intrinsic noise case. See for instance
  Ref.~\cite{Tonner-AP-05}, pp: 201.}

The mean-field approach conveyed by the network model has been
criticized on the basis that it neglects the coupling between local
density and global order produced by the motion of the
particles, \cite{Chate-comment}. This coupling can generate spacial
structures such as clusters of particles \cite{Huepe-PRL-04}, whereas
in the mean-field theory one assumes that the system is spatially
homogeneous. While this is true, is not clear how the coupling between
order and density in the Vicsek model changes the nature of the
phase transition.

\subsection{Hydrodynamic models}

For relatively large densities, one can imagine a flock as a continuum
fluid in which neighboring fluid cells tend to align their
momenta. This was the approach taken by Toner and
Tu, \cite{Toner-PRL-95}-\cite{Tonner-AP-05}, who constructed a
phenomenological hydrodynamic equation which has the same symmetries
as the Vicsek model. Specifically, this equation includes isotropic
short-range interactions, a self-propelled term, a stochastic driving
force, and does not conserve momentum. Together with mass
conservation, the system of hydrodynamic equations studied by Toner
and Tu is
\begin{eqnarray}
 \partial_t\vec{v} + \lambda_1(\vec{v}\cdot \nabla) \vec{v} +
 \lambda_2(\nabla \cdot \vec{v})\vec{v} + \lambda_3\nabla(|\vec{v}|^2)
 &=& \alpha\vec{v}-\beta|\vec{v}|^2\vec{v} - \nabla P +
 D_B\nabla(\nabla\cdot \vec{v}) + D_T \nabla^2 \vec{v} +
 D_2(\vec{v}\cdot\nabla)^2\vec{v} + \vec{f}, \label{eq:toner} \\ 
 P(\rho) &=& \sum_{n=1}^\infty\sigma_n(\rho-\rho_0)^n, \\ 
\partial_t \rho + \nabla\cdot(\vec{v}\rho) &=& 0,
\end{eqnarray}
where $\vec{v}$ is the velocity field, $\rho$ the density, $P$ the
pressure, $\sigma_n$ are the coefficients in the pressure expansion,
and $\vec{f}$ is a random Gaussian white noise which is equivalent to
the noise term in the Vicsek model. Through a
renormalization analysis of the above equations, Toner and Tu found
scaling laws and relationships between different scaling
exponents. Although the authors do not explicitly mention the nature
of the phase transition, the characterization they present in terms of
scaling laws and scaling exponents is, again, not inconsistent with a
continuous phase transition in 2D.

Later, Bertin et al. derived an analogous hydrodynamic equation from
microscopic principles incorporating only binary ``collisions'' of
self-propelled particles with intrinsic noise into a Boltzmann
equation, \cite{Bertin-PRE-06}. One of the advantages of this approach
is that all the phenomenological coefficients appearing in
Eq.~(\ref{eq:toner}), (the $\lambda$'s, $\alpha$, $\beta$, etc.), can
be explicitly expressed in terms of microscopic quantities. The
authors then show that the hydrodynamic equation has a homogeneous
non-zero stationary solution that appears continuously as the density
and the noise are varied. However, linear stability analysis shows
that for this model, this homogeneous solution is unstable and
therefore, the authors conclude, more complicated structures should
appear in the system (such as bands or stripes, see
Sec.~\ref{sec:artifact}).

In any case, the evidence collected so far with hydrodynamic models
of swarming, formulated along the same lines as the Vicsek model,
while is not inconsistent with a second order phase transition,
it has not been able either to shed light on the
nature of the transition.

\subsection{Models based on alignment forces}

In a different type of swarming models the interactions between the
particles are modeled through forces that tend to align the
velocities of the interacting particles. Newton equations of motion
are then solved for the entire system. Along these lines, Peruani et
al. analyzed a system of over-damped particles whose dynamics is
determined by the equations of motion \cite{Peruani-EPJ-08}
\begin{subequations}
\label{eq:peruani}
\begin{eqnarray}
\frac{d\vec{x}_m(t)}{dt} &=& v_0 e^{i\theta_m(t)}
\\ \frac{d\theta_m(t)}{dt} &=& -\gamma\frac{\partial
  U_m(\vec{x}_n,\theta_n)}{\partial\theta_m} + \eta_m(t)
\end{eqnarray}
\end{subequations}
where $\gamma$ is a relaxation constant, $\eta_m(t)$ is a Gaussian
white noise (similar to the intrinsic noise in Vicsek's
model), and the potential interaction energy $U_m(\vec{x}_n,\theta_n)$
is given by
\[
 U_m(\vec{x}_n,\theta_n) = \sum_{|\vec{x}_m-\vec{x_n}|< r_0}
 \cos(\theta_m-\theta_n).
\]
where $r_0$ is the radius of the interaction vicinity. In the limiting
case $\gamma\to\infty$ of very fast angular relaxation, the above
model is completely equivalent to the Vicsek
model, \cite{Peruani-EPJ-08}.

Analogously, inspired by the migration of tissue cells (keratocytes),
Szab\'o et al. formulated a model of over-damped swarming particles for
which the dynamic equations of motion are \cite{Szabo-PRE-06}
\begin{subequations}
 \label{eq:szabo}
\begin{eqnarray}
\vec{v}_m(t) &=&\frac{d\vec{x}_m(t)}{dt} = v_0 \hat{n}_m(t) +
\mu\sum_{n=1}^N \vec{F}(\vec{x}_m,\vec{x}_n)\\ \frac{d\theta_m(t)}{dt}
&=&
\frac{1}{\tau}\arcsin\left(\left|\hat{n}_m(t)\times\frac{\vec{v}_m(t)}{|\vec{v}_m(t)|}\right|\right)
+ \eta_m(t)
\end{eqnarray}
\end{subequations}
where $\theta_m$ is the angle of the unit vector $\hat{n}_m$, and
$\eta_m(t)$ is again a Gaussian white noise. The
interaction force $\vec{F}(\vec{x}_m,\vec{x}_n)$ depends piece-wise
linearly on the distance $d_{mn}=|\vec{x}_m-\vec{x}_n|$. It is
repulsive if $d_{mn} < R_0$, attractive if $r_0\leq d_{mn} \leq R_1$,
and vanishes if $R_1 < d_{mn}$. The explicit form of the force is
\[
 \vec{F}(\vec{x}_m,\vec{x}_n) = \hat{e}_{mn}\times\left\{
\begin{array}{lcl}
F_r\frac{d_{mn}-R_0}{R_0} & \mbox{ if } & d_{mn} < R_0 \\
F_a\frac{d_{mn}-R_0}{R_1 - R_0} & \mbox{ if } & R_0 \leq d_{mn} \leq R_1\\
0 & \mbox{ if } & R_1 < d_{mn}
\end{array}\right.
\]
where $\hat{e}_{mn}$ is the unit vector pointing from $\vec{x}_n$ to
$\vec{x}_m$, and $F_r$ and $F_a$ are the magnitudes of maximum
repulsion and maximum attraction, respectively.

\begin{figure}[t]
\centerline{\psfig{file=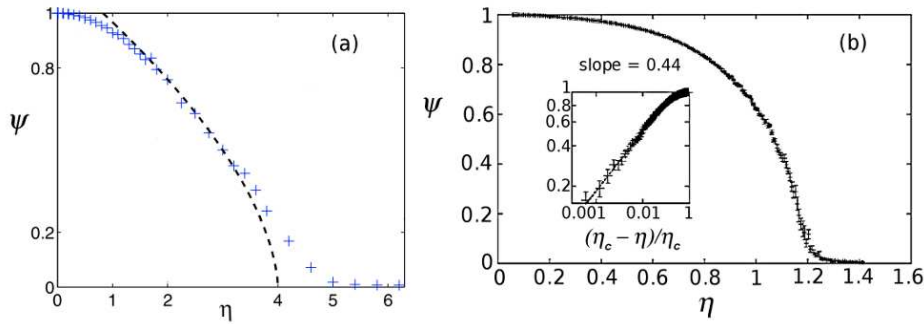,width=5in}}
\vspace*{8pt}
\caption{Phase transitions in swarming models based on alignment
  forces with intrinsic noise. (a) Order parameter as a function of
  the noise intensity for the model given in
  Eqs.~(\ref{eq:peruani}). The dashed curve is the mean-field
  prediction whereas the symbols correspond to numerical
  simulations. (b) Phase transition for the model given in
  Eqs.~(\ref{eq:szabo}). The data were obtained from numerical
  simulations. In both cases the phase transition is continuous. The
  figures (a) and (b) were taken (with small modifications to make the
  labels clearer) from Refs.~\cite{Peruani-EPJ-08} and \cite{Szabo-PRE-06}, respectively.}
\label{fig:8}
\end{figure}

In the two models described above, the forces between neighboring
particles tend to align their velocities if the particles are withing
the specified region, and in both models the particles are subjected
to intrinsic noise. Interestingly, numerical simulations show that the
states of collective order in these two models appear continuously
through a second order phase transition, as Fig.~\ref{fig:8}
illustrates.

\section{The Discontinuous Phase Transition}
\label{sec:discontinuous}

\begin{figure}[h]
\centerline{\psfig{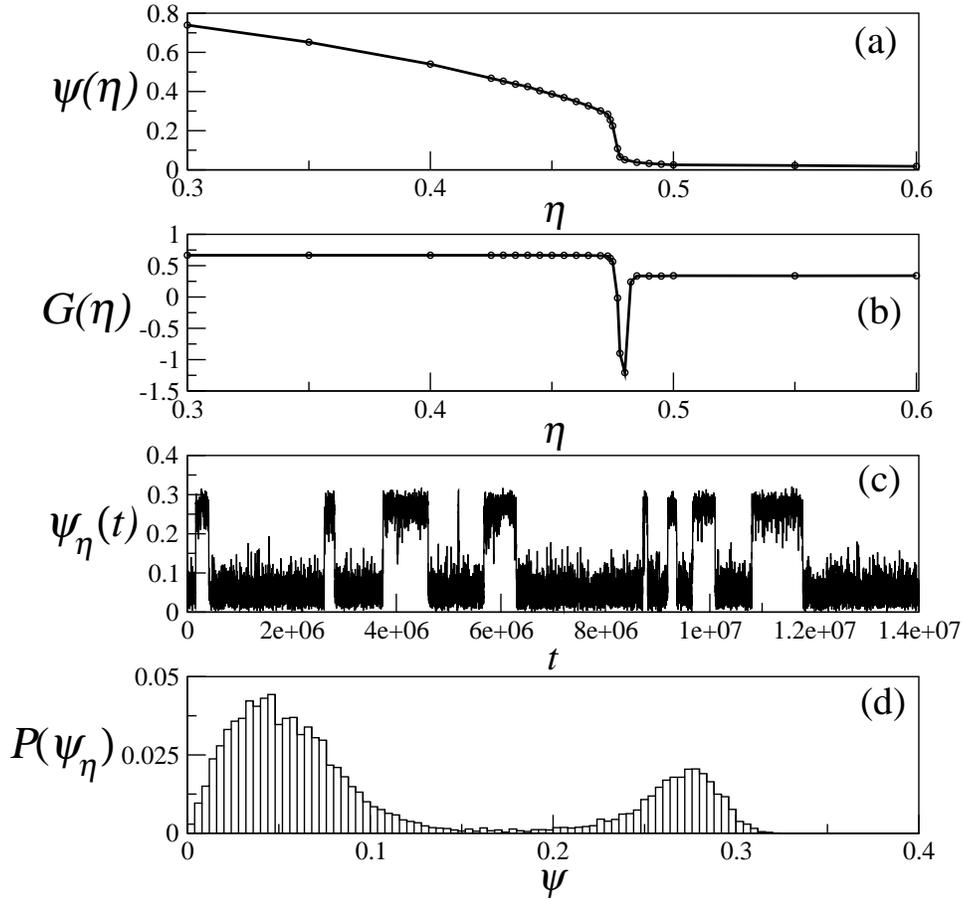}}
\vspace*{8pt}
\caption{\small\small Evidence of the discontinuous character of the phase
  transition in the Vicsek model with intrinsic noise, for as system
  with $N = 131072$ particles, $L=256$, and $l = 0.5$. (a) Order
  parameter $\psi$ as a function of the noise intensity $\eta$. In
  this case the order parameter decays to zero discontinuously around
  the critical value $\eta_c\approx0.478$.  (b) Binder cumulant $G$ as
  a function of $\eta$. Note the sharp minimum of the Binder cumulant
  close to the phase transition. (c) Time series of the instantaneous
  order parameter $\psi_\eta(t)$ for $\eta=0.475$ (very close to the
  phase transition). This time series exhibits bistability, which is
  characteristic of phase coexistence in a first-order phase
  transition. (d) Probability distribution function
  $P(\psi_\eta)$ of the instantaneous order parameter
  $\psi_\eta(t)$. This PDF is bimodal because of the bistability of
  the time series. In (a) and (b) each data point is the average of
  $\psi_\eta(t)$ over $2\times10^7$ time steps. }
\label{fig:9}
\end{figure}

The first evidence for the discontinuous character of the phase
transition in the Vicsek model with intrinsic noise appeared in
Ref.~\cite{Chate-PRL-04}, and it was later reinforced in
Ref.~\cite{Chate-PRE-08}. In that work the authors studied very
large systems in the high-velocity regime 
$l\geq 0.5$, \footnote{This regime may be biologically relevant for swarms of
  insects such as locusts or cannibal crickets, which interact mainly
  through contact forces, \cite{Buhl-Science-06,Simpson-PNAS-06}. In
  these systems, one can assume that in each jump, the insects travel a
  distance comparable to the size of their interaction vicinity.}.
Fig.~\ref{fig:9} summarizes this evidence for a large
system, with $N=131072$ particles, $L=256$, and $l=0.5$. This system
is similar to one of the systems used by Gr\'egoire and Chat\'e in
Ref.~\cite{Chate-PRL-04} to demonstrate the discontinuity of the
phase transition with intrinsic noise. It is clear from
Fig.~\ref{fig:9}a that the order parameter undergoes an
apparently discontinuous phase transition around a critical value
$\eta_c=0.478$. Such a discontinuity is confirmed by the Binder
cumulant $G(\eta)$, defined as
\begin{equation}
G(\eta) = 1-\frac{\langle \psi_\eta^4\rangle}{3\langle \psi_\eta^2\rangle^2}.
\label{eq:binder-cumulant} 
\end{equation}

The Binder cumulant, which is closely related to the Kurtosis, can
intuitively be interpreted as a measure of how much a probability
distribution function (PDF) differs from a Gaussian, particularly
concerning the ``peakedness'' of the distribution. It was introduced
by Binder as a means to determine numerically the nature of a phase
transition, \cite{Binder-RPP-87,Tsai-BJP-98}. The idea behind this
measure is that in a first-order phase transition, typically two
different metastable phases coexist. Throughout time, the system
transits between these two phases and consequently the time series of
the order parameter shows a bistable behavior, as it is illustrated in
Fig.~\ref{fig:9}c. The PDF $P(\psi_\eta)$ of this time
series is then bimodal, reflecting the coexistence of the two
different phases and the transit of the system between them (see
Fig.~\ref{fig:9}d). Because of this bimodality of
$P(\psi_\eta)$, which occurs only close to the phase transition, the
Binder cumulant has a sharp minimum, as it is actually observed in
Fig.~\ref{fig:9}b. On the other hand, far from the phase
transition, either in the ordered regime or in the disordered one,
$P(\psi_\eta)$ is unimodal and rather similar to a Gaussian, for which
the Binder cumulant is constant. It is only close to a discontinuous
phase transition that $P(\psi_\eta)$ becomes bimodal, thus
deviating from a Gaussian, and the
Binder cumulant $G(\eta)$ exhibits a sharp minimum. This is the
strongest evidence presented by Gr\'egoire and Chat\'e in favor of a
discontinuous phase transition in the Vicsek model with intrinsic
noise, \cite{Chate-PRL-04,Chate-PRE-08}. They also analyzed variations
of this model, for instance by introducing cohesion or substituting
the intrinsic noise by extrinsic noise. However, the mean field
network models presented in Sec.~\ref{sec:networks} show that the
results obtained for one type of noise cannot be extrapolated to a
different type of noise. Thus, in this section we will focus on the
intrinsic noise case.

It should be emphasized that the discontinuous behavior displayed in
Fig.~\ref{fig:9} is observed only in the high velocity
regime. Indeed, it is apparent from Fig.~\ref{fig:4} that in the
low-velocity regime $l=0.03$, the order parameter $\psi(\eta)$ does
not exhibit any discontinuity even for a large system with $N=131044$
and $L=181$, (comparable in size to the one used in
Fig.~\ref{fig:9}). It could be possible that the phase
transition changes from continuous to discontinuous across the
parameter space. However, the authors of Refs.~\cite{Chate-PRL-04}
and \cite{Chate-PRE-08} claim that this does not happen. They
state that all phase transitions occur in the thermodynamic limit, and
if we were able to explore very large systems, then the phase
transition in the Vicsek model with intrinsic noise would turn out to
be discontinuous for any values of the parameters, \cite{Chate-comment}.

Another important aspect of the discontinuous phase transition shown
in Fig.~\ref{fig:9}a is the absence of hysteresis. Indeed,
our numerical simulations show that for this system it is irrelevant
to start out the dynamics from ordered or from disordered states,
because in either case the curve displayed in
Fig.~\ref{fig:9}a is obtained. This is in marked contrast
with the hysteresis commonly observed in usual first-order phase
transitions, as the one depicted in Fig.~\ref{fig:7}b,
which corresponds to the extrinsic noise dynamics and therefore is
undoubtedly first order and in accord with the mean-field analytical
results (see Sec.~\ref{sec:networks}). In the next section we present
evidence indicating that the discontinuity of the phase transition
observed in Fig.~\ref{fig:9} may in fact be an artifact of
the periodic boundary conditions, as first proposed by Nagy et
al, \cite{Nagy-PA-07}.

\section{The Effect of the Boundary Conditions}
\label{sec:artifact}

\begin{figure}[ht]
\centerline{\psfig{file=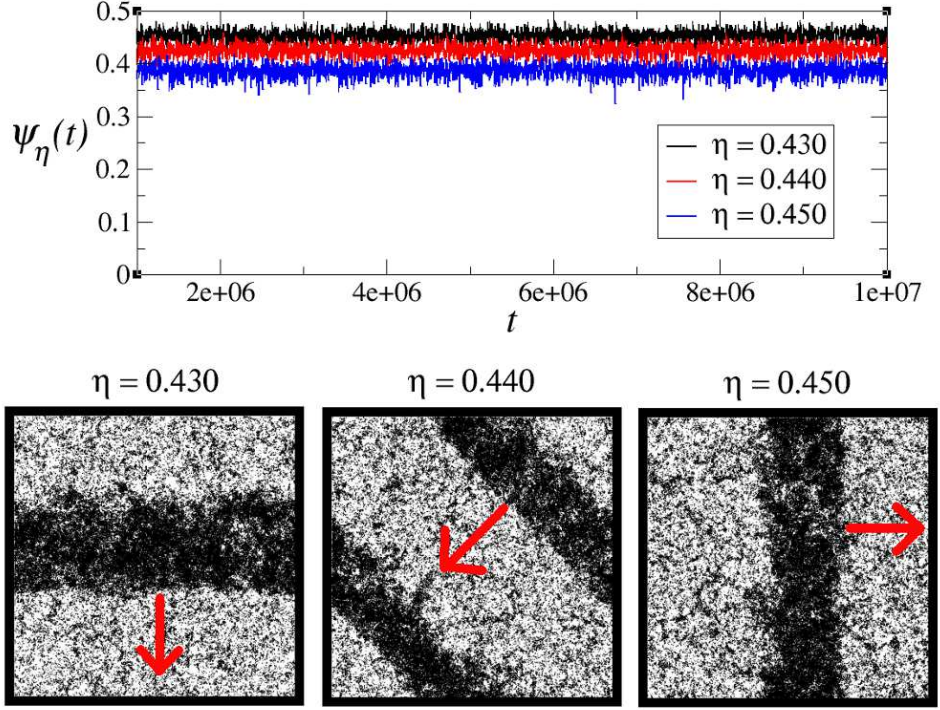,width=5in}}
\vspace*{8pt}
\caption{Traveling bands in the Vicsek model with intrinsic
  noise. The top panel shows the time series of the order parameter
  for three different values of the noise intensity $\eta$. The images
  at the bottom are snapshots of the system corresponding to each of
  the three values of $\eta$ used. The arrows indicate the direction of
  motion of the bands.}
\label{fig:10}
\end{figure}

In a recent paper \cite{Nagy-PA-07}, Nagy, Daruka and Vicsek  pointed
out the formation of traveling bands in the high velocity regime. In
Fig.~\ref{fig:10} we show these bands for three different
values of the noise intensity $\eta$ and for the same system used to
compute the phase transition in Fig.~\ref{fig:9}. Note that
the three values of $\eta$ in Fig.~\ref{fig:10},
($\eta=0.430$, $\eta=0.440$ and $\eta=0.450$), correspond to the
ordered regime, quite far from the transition, which occurs around
$\eta_c\approx0.478$. Interestingly, the bands always wrap around
tightly in a periodic direction, which for simple shapes is usually
either perpendicular or parallel to a boundary of the periodic box, or
along a diagonal.  In the ordered regime, the bands are very
stable. Once they form, they may travel for extremely long times in the
same direction (as shown in Fig.~\ref{fig:10},
the motion of the bands, indicated by the arrow, is perpendicular to themselves). In
Ref.~\cite{Nagy-PA-07} the authors also use hexagonal boxes with
periodic boundary conditions, and the bands again always form parallel
to the boundaries of the hexagon or to its diagonals.

\begin{figure}[t]
\centerline{\psfig{file=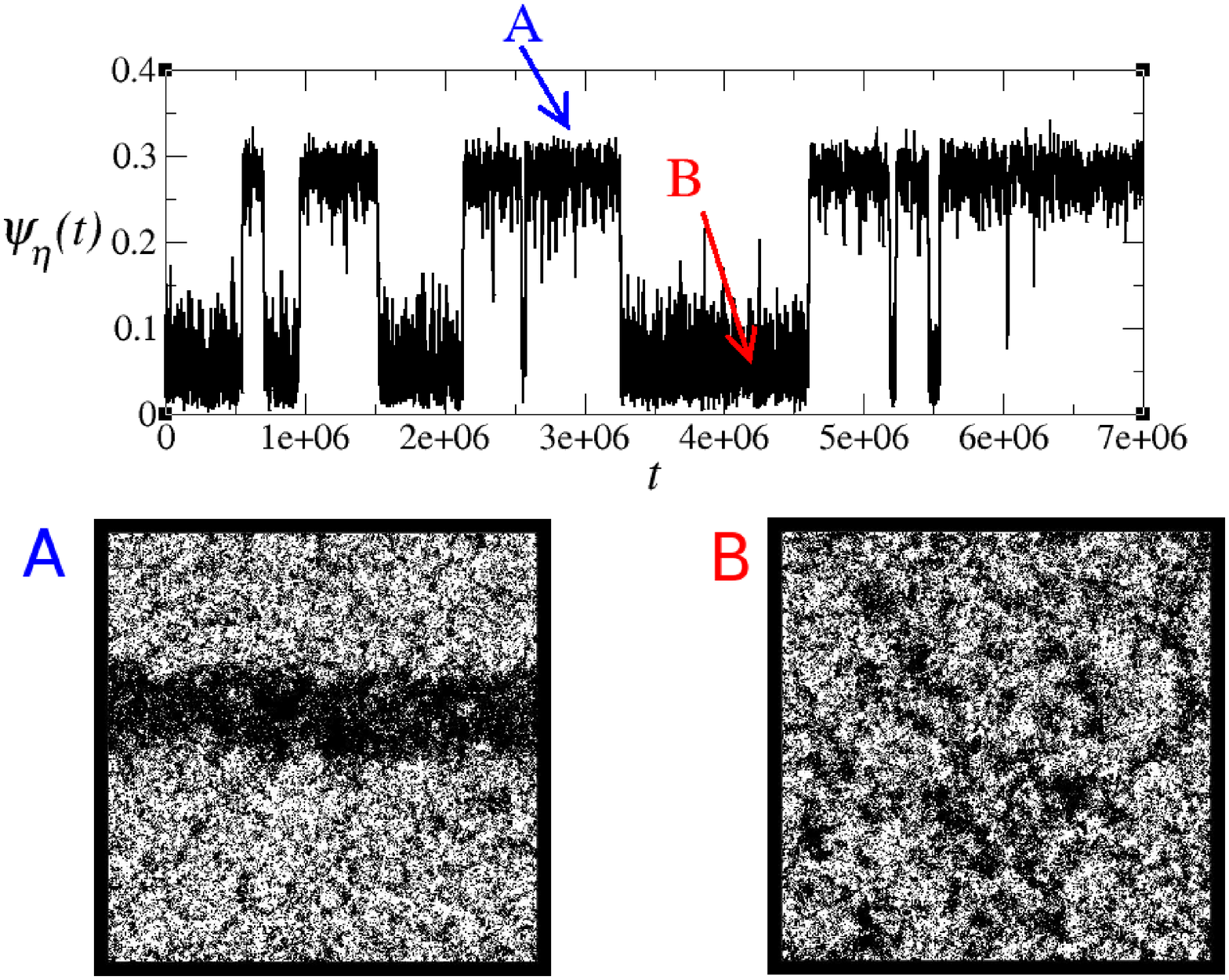,width=5in}}
\vspace*{8pt}
\caption{Bistable time series of the order parameter at a noise
  intensity $\eta=0.475$ for the same system as in
  Fig.~\ref{fig:9}. This value is very close to the
  critical value $\eta_c=0.478$. The bottom figures are snapshots of
  the system at the times indicated by the points A and B in the time
  series. The apparent bistability of the order parameter indicates
  the formation and destruction of the bands. }
\label{fig:11}
\end{figure}

When the value of $\eta$ approaches the critical value $\eta_c$, the
bands become unstable: They appear and disappear intermittently
throughout the temporal evolution of the system.  Fig.~\ref{fig:11}
shows the time series of the order parameter for the same system used
in Fig.~\ref{fig:9}, at a noise intensity $\eta=0.475$,
i.e. very close to the discontinuity. The time series shows that the
order parameter ``jumps'' between two rather different values. These
jumps give rise to the bimodality of the order parameter, which,
according to the interpretation presented in the previous section,
indicates phase coexistence. A close inspection of the spatial
structure of the system reveals that such bistability is produced by
the creation and destruction of the traveling bands, as the two
snapshots in Fig.~\ref{fig:11} indicate. Thus, as time
passes, the system transits from spatially homogeneous states to
states characterized by traveling bands. The authors of
Ref.~\cite{Chate-PRE-08} interpreted these bands as one of the
two metastable phases, which coexists with a more or less homogeneous
background of diffusive particles (the other metastable phase).
However, Vicsek and his group retorted that the formation of
traveling bands is not an intrinsic property of the system, but,
rather, an artifact of the periodic boundary 
conditions.\footnote{It is actually rather hard to envision a stationary scenario
  consisting of bands in an infinite system with no boundaries. If the
  bands were stable, one could expect the formation of several
  indefinitely long bands in such a system. Such bands would collide
  with each other as there is no reason to expect all the bands to be
  parallel to each other and to travel in the same direction. 
  Upon collision, non parallel bands merge or break apart, and it is not even clear that such system 
  will ever reach a stationary state.}

\begin{figure}[t]
\centerline{\psfig{file=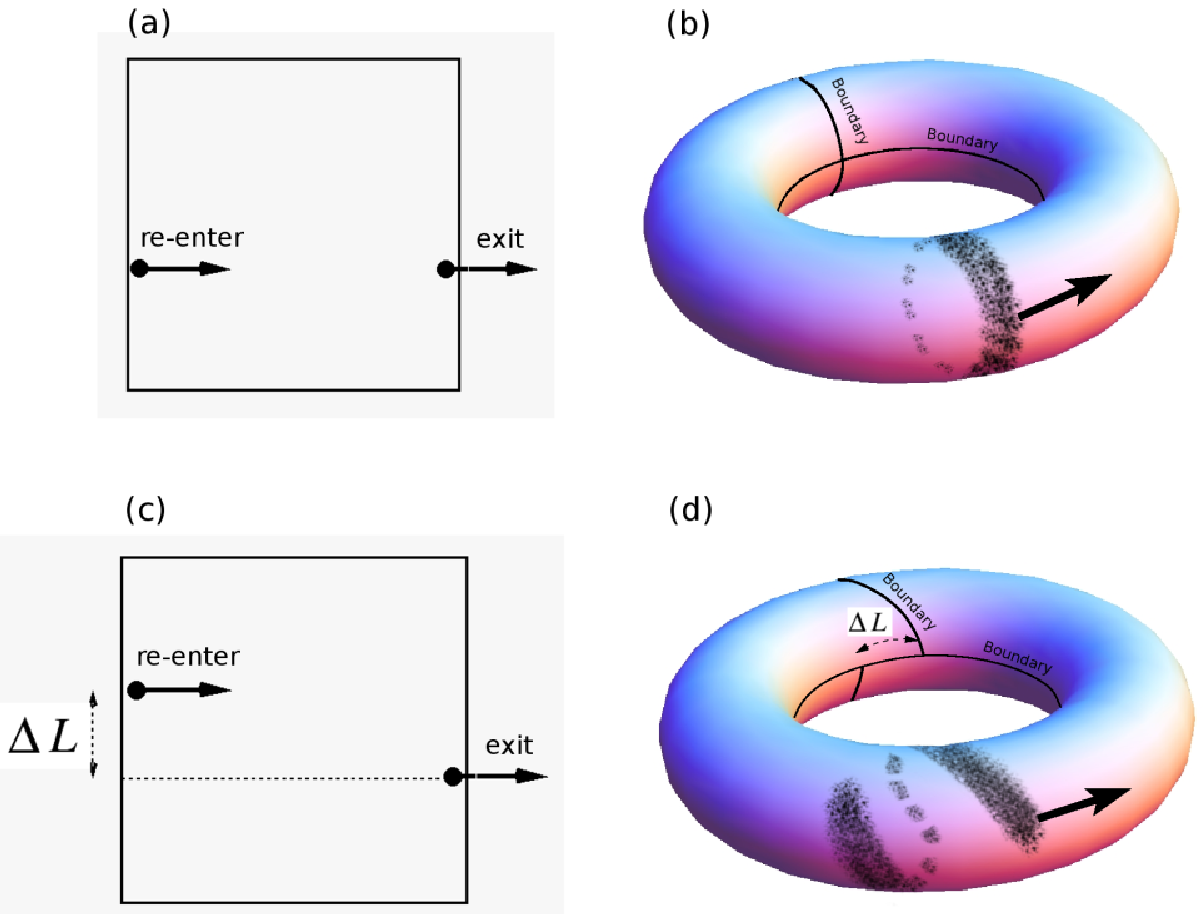,width=5in}}
\vspace*{8pt}
\caption{(a) In the original Vicsek model one implements periodic
  boundary conditions: If a particle exits through one side of the
  box, it re-enters through the opposite side at a specular
  position. (b) The periodic boundary conditions induce a toroidal
  topology that favors the formation of bands (closed rings) moving on
  the torus' surface. (c) Shift $\Delta L$ in the boundary
  conditions. The shift is implemented in the four sides of the box,
  which breaks the toroidal symmetry of the dynamics and makes the
  closed ring structures unstable, as it is illustrated in (d).  }
\label{fig:12}
\end{figure}

It is a very common resource in numerical simulations of physical
systems to use periodic boundary conditions as way to represent an
infinite homogeneous space and mitigate boundary effects. In Vicsek's
model the particles move in a box of sides $L$. If a given particle
exits the box through one of the sides, it is re-injected into the
system through the opposite side at a specular position (see
Fig.~\ref{fig:12}a). Under these circumstances, the real
geometry of the box is that of a torus, and the motion of the
particles takes place on its surface. The toroidal topology induced by
the periodic boundary conditions clearly favors the formation of
bands, which in fact are closed stable rings moving on the surface of
the torus, as is illustrated in Fig.~\ref{fig:12}b. However,
there is no reason whatsoever to re-inject the particles at specular
positions of the box. Rather, it is possible to re-inject the
particles at positions that are \emph{shifted} a distance $\Delta L$
with respect to the specular position, as is done in
Fig.~\ref{fig:12}c. By doing so, the toroidal topology is
broken and the formation of bands (closed rings) is not longer
stabilized by (toroidal) periodic boundary conditions. The idea is
schematically illustrated in Fig.~\ref{fig:12}c, although the
real situation is more complicated because the shift $\Delta L$ is
implemented in the four sides of the box (see the Appendix). 
This shift, which we will
refer to as the \emph{boundary shift}, does not affect the homogeneity
of the system, as all the particles are subjected to it when they hit
the boundary. Interestingly, when the shift is implemented, the
discontinuity of the phase transition reported in
Refs.~\cite{Chate-PRL-04} and \cite{Chate-PRE-08}, and
reproduced here in Fig.~\ref{fig:9}, seems to disappear.

\begin{figure}[t]
\centerline{\psfig{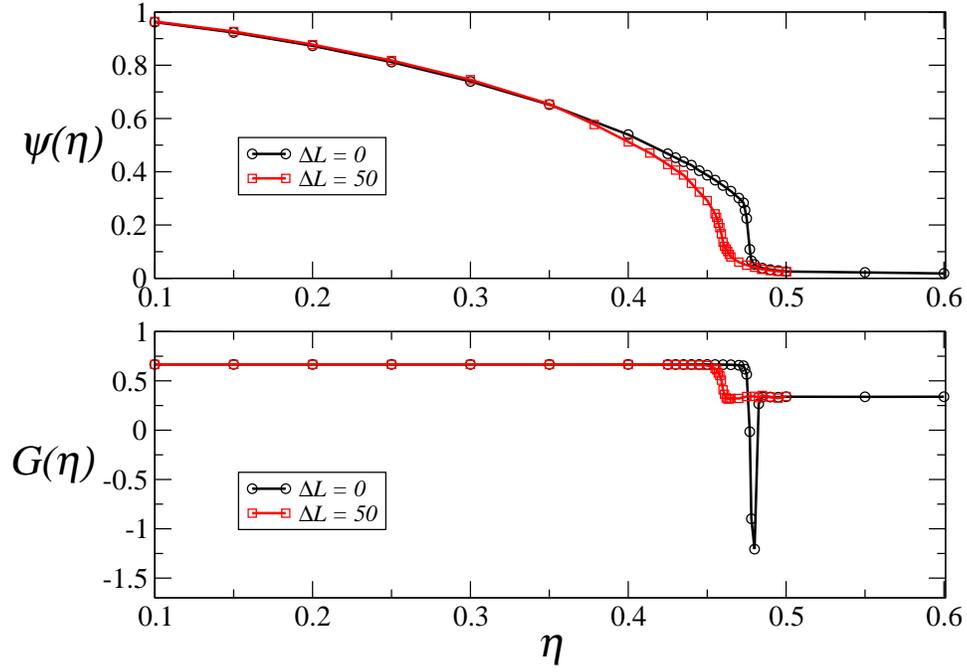}}
\vspace*{8pt}
\caption{Effect of the boundary shift $\Delta L$ on the phase
  transition in the Vicsek model with intrinsic noise. (a) Order
  parameter as a function of the noise intensity for a system with no
  shift (black curve with circles), and for a system with a boundary
  shift $\Delta L = 0.2L$ (red curve with squares). Note that the
  apparent discontinuity of the phase transition for the system with
  periodic boundary conditions disappears when the shift
  is introduced. (b) Binder cumulant corresponding to the two curves
  displayed in (a). The sharp negative peak disappears for the system
  with non-zero boundary shift. The curves were computed for systems
  with $N=131072$, $L=256$, and $l=0.5$. Every data point is the
  average of $\psi_\eta(t)$ over $2\times10^7$ time-steps. }
\label{fig:13}
\end{figure}

\begin{figure}[t]
\centerline{\psfig{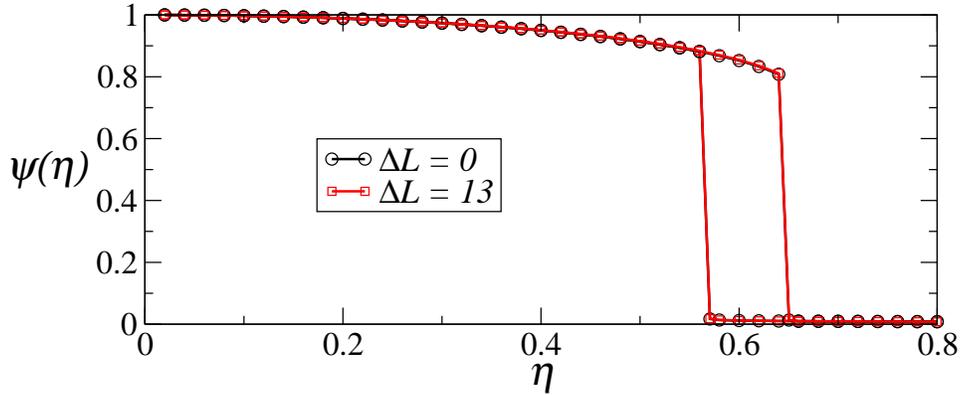}}
\vspace*{8pt}
\caption{Phase transition in the Vicsek model with extrinsic noise for
  systems with $N=80000$, $L = 64$, and $l = 0.5$. The black circles
  correspond to fully periodic boundary conditions, i.e. $\Delta L =
  0$, whereas the squares correspond to a boundary shift $\Delta L =
  0.2L$. It is clear that in this case the boundary shift does not
  change the phase transition, which is undoubtedly discontinuous.  }
\label{fig:14}
\end{figure}

Fig.~\ref{fig:13}a shows the phase transition for the same
system as the one used in Fig.~\ref{fig:9}, but this time
comparing the two cases: With fully periodic boundary conditions,
i.e. $\Delta L = 0$ (black curve with circles), and with a non-zero
boundary shift $\Delta L = 0.2L$ (red curve with squares). In
Fig.~\ref{fig:13}b the corresponding Binder cumulants are
reported. We choose $\Delta L = 0.2L$ because this is approximately
the width of the bands close to the transition (see snapshot A in
Fig.~\ref{fig:11}). It is clear from
Fig.~\ref{fig:13} that the introduction of the boundary shift
eliminates the discontinuity of the phase transition, making it
continuous. Note that the sharp peak in the Binder cumulant also
disappears. It could be argued, then, that if the phase transition was
intrinsically discontinuous, the introduction of the boundary shift
should not change its character. This is indeed what happens in the
Vicsek model with extrinsic noise with a boundary shift $\Delta L =
0.2L$, as illustrated in Fig.~\ref{fig:14}. In such case,
of course, the introduction of the boundary shift does not change the
appearance of the phase transition. The above results give support to
the claim that the apparent discontinuous character of the phase
transition in the intrinsic noise case may indeed be an artifact of
the topology induced by the periodic boundary conditions, as Vicsek
and his colleagues had already suggested, \cite{Nagy-PA-07}.

\begin{figure}[t]
\centerline{\psfig{file=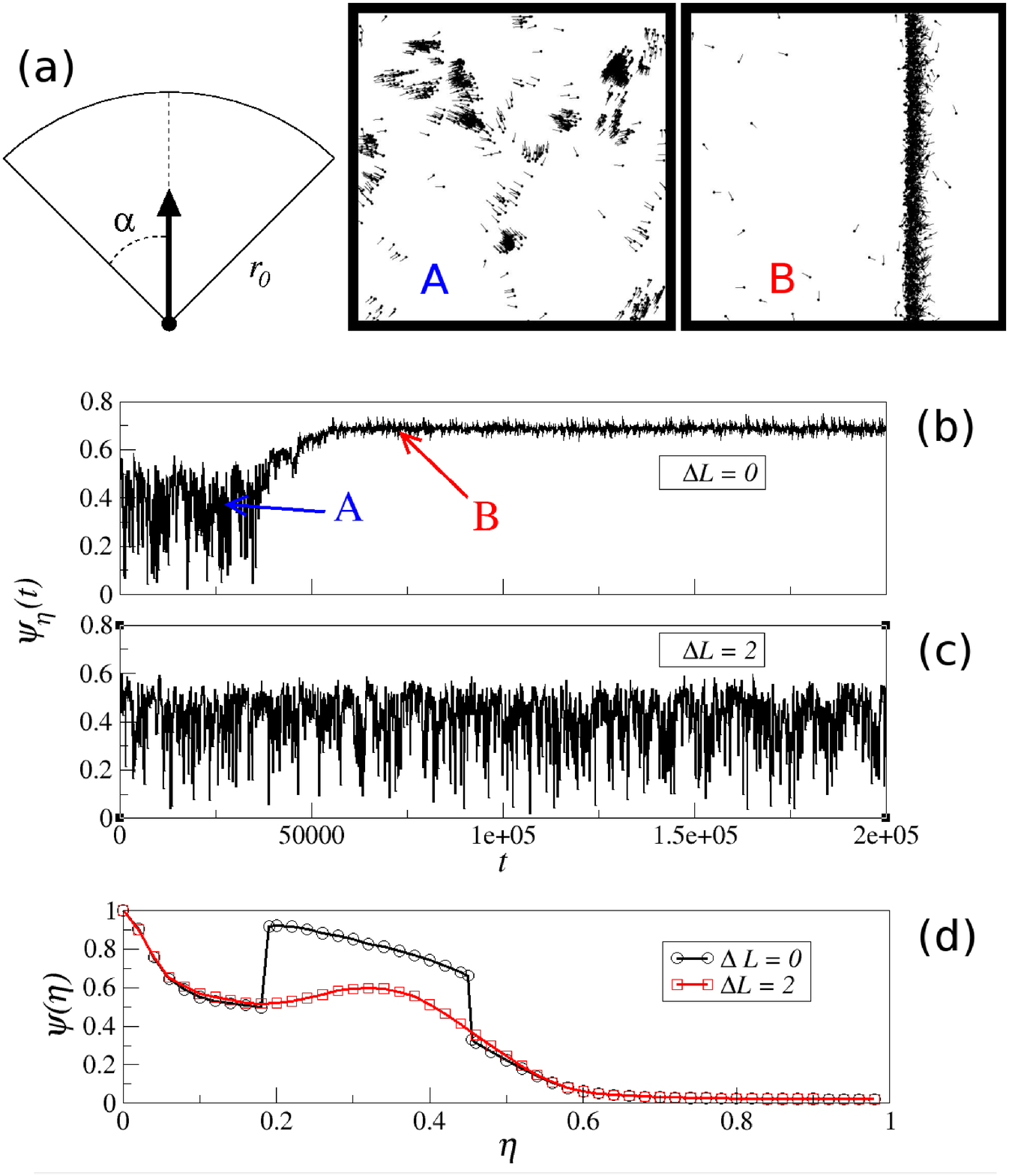,width=5in}}
\vspace*{8pt}
\caption{Phase transition in a variation of the Vicsek model with
  $N=1600$, $L=20$ and $l = 0.03$. (a) The interaction vicinity is a
  circular cone of radius $r_0=1$ and angular amplitude $2\alpha$,
  centered at the particle's velocity. (b) Time series of the order
  parameter for $\alpha = 0.4\pi$ and $\eta = 0.44$, with fully
  periodic boundary conditions ($\Delta L = 0$). Note that after a
  transient time, the particles aggregate in a very dense band
  parallel to the boundary. The snapshots A and B correspond to the
  points signaled with the arrows in the time series. (c) The same as
  in the previous panel but now with a boundary shift $\Delta L =
  0.1L$. (d) Phase transition diagram with (red curve) and without
  (black curve) boundary shift. Note the discontinuities produced by
  the periodic boundary conditions.}
\label{fig:15}
\end{figure}

A more dramatic effect of the periodic boundary conditions on the
phase transition can be observed in a simple variation of the Vicsek
model, for which it is not necessary to analyze very large systems. In
this variation, the particles have a ``vision angle''. In other words,
the interaction region of each particle is no longer a full circle
surrounding it, but only a circular cone of amplitude $2\alpha$
centered at the particle's velocity, as illustrated in
Fig.~\ref{fig:15}a. All the other aspects of the interaction
are as in Vicsek's model, only the form of the interaction vicinity is
changed. In what follows we consider such a system with intrinsic
noise and $N=1600$, $L=20$, $l = 0.03$ (low-velocity regime), and $\alpha = 0.4\pi$. For
periodic boundary conditions and some values of the intrinsic noise,
after a transient time the particles form very dense bands, as shown
in Fig.~\ref{fig:15}b and the corresponding
snapshots. Actually, as time goes by, the system transits back and
forth, from configurations like the one depicted in snapshot A, to
configurations with bands, as the one shown in snapshot B. Those bands
produce \emph{two} discontinuities in the curve of the order parameter,
which is plotted in Fig.~\ref{fig:15}d (black curve with
circles). Interestingly, these discontinuities do not occur close
to the phase transition (which in this case seems to happen around
$\eta_c \approx 0.6$), but well into the ordered phase. On the other
hand, Fig.~\ref{fig:15}c shows the time series of the order
parameter when a boundary shift $\Delta L = 0.1L$ is introduced. Note
that in this case the order parameter does not reflect the transit of
the system between any two phases. This is because with the boundary
shift the bands no longer form and the two discontinuities in the
phase transition curve disappear, as is apparent from
Fig.~\ref{fig:15}d (red curve with squares).

This variation of the Vicsek model clearly demonstrates that the
periodic boundary conditions can have a dramatic effect on the system
dynamics, producing artificial discontinuities in the order parameter
that would not necessarily be present in an infinite system with
no boundaries.

\section{Summary}
\label{sec:summary}

In this work we have presented much of the evidence offered in support of two distinct points of view, at odds with each other, about the nature of the phase transition in the Vicsek model of self-propelled particles with intrinsic noise. On the one hand, there is theoretical and numerical evidence indicating that such a phase transition is continuous. This evidence comes from several fronts: From the numerical side, finite-size scaling analysis and dynamic scaling analysis performed for systems operating in the low-velocity regime, have yielded results consistent with a continuous phase transition. These numerical techniques show scaling behavior of the order parameter with the size of the system and with the density, for systems as large as the computer capacity allows one to analyze. Additionally, the hyperscaling relationship between the critical exponents has been confirmed within numerical accuracy. This scaling behavior is consistent with what is known for second order phase transitions in equilibrium systems. From the theoretical side, analytic calculations unambiguously show that the mean-field version of the Vicsek model with intrinsic noise undergoes a second-order phase transition. Furthermore, swarming models based on alignment forces, (rather than on kinematic rules), also exhibit continuous phase transitions when the intrinsic noise is used. 

On the other hand, there is direct numerical evidence suggesting that the phase transition is discontinuous in very large systems operating in the high-velocity regime. In addition to a clear discontinuity in the phase transition curve for these large systems, this point of view is supported by an apparent bistability of the order parameter close to the phase transition, and the corresponding bimodality of its PDF signaled by a ``dip'' of the Binder cumulant. Such a scenario is characteristic of phase coexistence in first-order transitions. 

However, numerical computations performed for systems comparable in size to those used to demonstrate that the phase transition is discontinuous, suggest that the apparent discontinuity is an artifact of the toroidal topology induced by the periodic boundary conditions. For, when the boundary conditions are ``shifted'' in order to break the periodicity, and hence the toroidal symmetry, the discontinuity disappears. This effect is enhanced when the particles have a limited vision angle.

Determining the nature of the phase transition in the Vicsek model with intrinsic noise is important, among other things, for understanding the physical principles through which collective order emerges in swarms and flocks, and more generally, in systems out of equilibrium. Despite the efforts of many groups during the last decade, nobody has said the last word about the onset of collective order in swarming systems. This remains an open problem.

\section*{Acknowledgements}

We thank C. Huepe for useful comments and suggestions. This work was supported by CONACyT Grant P47836-F, and by PAPIIT-UNAM Grant IN112407-3. 

\appendix
\section{Boundary shift}

\begin{figure}[t]
\centerline{\psfig{file=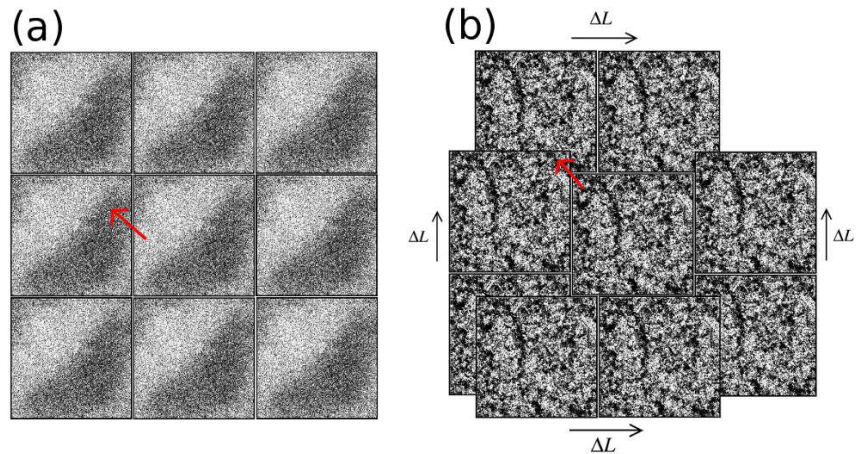,width=4.5in}}
\vspace*{8pt}
\caption{(a) A system with periodic boundary conditions. This is equivalent to making infinite copies of the central box and covering homogeneously the space with them. (b) A system with boundary shift. Note that the shifted boxes overlap at the corners of the central box.}
\label{fig:A1}
\end{figure}

In this appendix we explain how the boundary shift $\Delta L$ was implemented in our numerical simulations. Let us start with a system with periodic boundary conditions, as in Fig.~\ref{fig:12}a. Using periodic boundary conditions is equivalent to making infinite copies of the original system, and covering homogeneously all the space with these copies. This is illustrated in Fig.~\ref{fig:A1}a, where a central box has been ``cloned'' several times. A particle exiting the central box just re-enters into it at a position equivalent to the one it would have in the adjacent box, and so on. 

On the other hand, Fig.~\ref{fig:A1}b shows how the boundary shift $\Delta L$ is implemented. The copies of the central box have been shifted around it the same positive distance $\Delta L$ in both the horizontal and the vertical directions. In this case, the space is not cover homogeneously because the shifted boxes overlap, as it is clearly seen in Fig.~\ref{fig:A1}b. However, this overlap is important only for particles exiting the central box from the corners (as the one signaled with an arrow), because such a particle does not ``know'' what box to go to. This overlapping mixes the particles exiting from the corners in a way which is different from the natural motion of the particles located at the bulk of the system. Thus, the boundary shift $\Delta L$ introduces inhomogeneities in the ``tiling'' of the space. These inhomogeneities occur in regions of size $v_0\Delta t$ around the corners of the central box. Therefore, if $(v_0\Delta t)/L \ll 1$, as it has been the case in all the simulations carried out in this work, the inhomogeneities generated by the boundary shift should be negligible. In our simulations, when a particle exits the central box from one of the corners, it randomly chooses one of the two overlapping boxes to be re-injected into it.

\end{document}